\documentclass[twocolumn,showpacs,preprintnumbers,amsmath,amssymb]{revtex4}
\usepackage{graphicx}
\usepackage{dcolumn}
\usepackage{bm}
\begin{document}

\title{Precise calculation of the threshold of various directed
percolation models on a square lattice}
\author{Danyel J. B. Soares}
\author{Jos\'e S. Andrade Jr.}
\author{Hans J. Herrmann}
\altaffiliation[Also at ] {Institute for Computer Physics,\\
 University of Stuttgart\\
email: hans@ica1.uni-stuttgart.de }
 \affiliation{%
Departamento de F\'{i}sica, Universidade Federal do Cear\'{a}\\
60451-970 Fortaleza, Cear\'{a}, Brazil}

\date{\today}


\begin{abstract}
Using Monte Carlo simulations on different system sizes we determine
with high precision the critical thresholds of two families of
directed percolation models on a square lattice. The
thresholds decrease exponentially with the degree of connectivity. We
conjecture that $p_{c}$ decays exactly as the inverse of the coodination
number.
\end{abstract}

\maketitle


Directed percolation (DP) describes generically the dynamics
of adsorbing processes and has been applied to
epidemics, forest fires, surface catalysis, etc [1-6]
DP displays a phase transition as function of the
propagation probability $p$ between an ``adsorbing"
state and a percolating (or active) state at a
critical threshold $p_c$. The scaling properties of DP
have been known since over two decades.

The value of $p_c$ depends on the lattice and the rule
of connectivity. In 1+1 dimension the most studied
case is the tilted square lattice with
nearest neighbor connectivity (see Fig.~1) giving
$p_c=0.6447 \pm 0.0001$ \cite{Jen}. For many applications
it is of interest to consider a longer range of
connectivities. In particular also an infinite range
model with a probability decaying as a powerlaw with
distance has been studied \cite{Hin}. However, what is
astonishingly missing in the literature are studies
of propagation probabilities of intermediate range
although these are in practice the most common situations.
It is therefore the aim of the present
paper to calculate the percolation thresholds
of two families of models in 1+1 dimensions having
finite, varying range of interactions.

\begin{figure}
\begin{center}
 \includegraphics[width=0.4\textwidth]{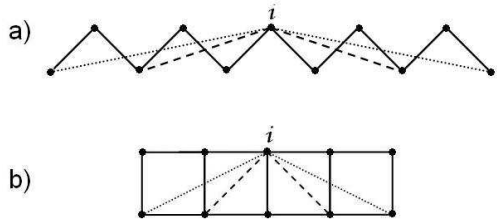} \
\end{center}
\caption[kurzform]{\label{Fig1} Schematic plot of the definitions
of (a) model I and (b) model II. The full lines correspond to
(a) $k= $ and (b) $k=1$, adding the dashed lines one
obtains (a) $k=4$ and (b) $k=3$ and adding the dotted lines
one obtains (a) $k=6$ and (b) $k=5$.}
\end{figure}

Model I is defined on a tilted square lattice as
shown in Fig.~1a. The connectivity $k$ is given
as the number of neighbors to which a site $i$ can be
connected in the row below. In model I the $k$ is
always even. Fig.~1a also shows the cases $k=2,4$ and 6.
Model II is defined on the standard square lattice
and in Fig.~1b we can see the cases $k=1,3$ and 5.
For this case, $k$ is always odd.

We studied the two models up to $k=15$ on lattices
of sizes ranging from $256\times20000$ till $2048\times20000$.
We simulated for each value of $k$ and $L$ the model
for different values of $p$ averaging over $2000$
initial configurations and monitored if the
system percolated or not. From the inflection point
of the histogram as function of $p$ we determined $p_c(L)$.
Finally we extrapolated to the thermodynamic limit $p_c$
through the expression
$$p_c(L)=p_c (1+a L^{-1/\nu})\eqno(1)$$.

\begin{figure}
\begin{center}
 \includegraphics[width=0.4\textwidth]{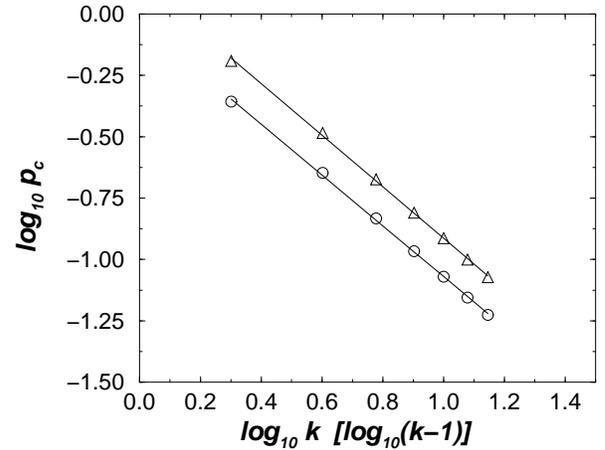} \
\end{center}
\caption[kurzform]{\label{Fig2} Double-logarithmic plot of the percolation
thresholds as function of $k$ for  model I (triangles) and of $k-1$ for
model II (circles).}
\end{figure}

In table~1 we show the obtained thresholds with
their estimated error bars and in Fig.~2 we
see these values in a double-logarithmic plot.
Since the slope is $-1\pm0.05$ we conclude that
$$p_c=\eta/k\eqno(2a)$$ with $\eta=1.3\pm0.1$ for model I
and  $$p_c=\eta/(k-1)\eqno(2b)$$ with $\eta=0.9\pm0.1$
for model II.

In summary, we have described in this short paper numerical evidence
that the percolation threshold of directed percolation
decays like $\eta/k$ as function of the coordination $k$.
This result is consistent with the dependence of the
percolation threshold on the coordination number on the
Cayley tree \cite{Ess} and we conjecture it to be exact.
 As a future work, it would be interesting to calculate 
the thresholds for higher dimensional lattices.

\begin{table}[htpb]

\begin{tabular}{|c|c|c|c|}

 \hline
 $k$&I&$k$&II\\
 \hline
 $2$&$0.6447\pm0.0002$&$3$&$0.4395\pm0.0003$\\
 \hline
 $4$&$0.3272\pm0.0002$&$5$&$0.2249\pm0.0003$\\
 \hline
 $6$&$0.2121\pm0.0003$&$7$&$0.1470\pm0.0002$\\
 \hline
 $8$&$0.1553\pm0.0003$&$9$&$0.1081\pm0.0002$\\
 \hline
 $10$&$0.1220\pm0.0002$&$11$&$0.0851\pm0.0002$\\
 \hline
 $12$&$0.0999\pm0.0002$&$13$&$0.0701\pm0.0002$\\
 \hline
 $14$&$0.0846\pm0.0003$&$15$&$0.0549\pm0.0002$\\
\hline
\end{tabular}

 \caption{Percolation threshold of directed percolation to model I and model II.}
 \end{table}

\bibliographystyle{prsty}

\end{document}